\documentclass{PoS}
\usepackage{bm}
\usepackage{epsfig,multicol}

\def\PTP{Prog. Theor. Phys.(Kyoto)}

\def\NPB{{Nucl. Phys.} {\bf B}}
\def\PLB{{Phys. Lett.} B}

\def\PRD{{Phys. Rev.} D}

\newcommand{\Slash}[1]{\ooalign{\hfil/\hfil\crcr$#1$}}

\title{Self-dual gauge fields, \\domain wall fermion zero modes\\
and the Kugo-Ojima confinement criterion}

\ShortTitle{Self-dual gauge fields and domain wall fermion zero modes}

\author{\speaker{Sadataka Furui}%
         \\
        School of Sci.and Engr., Teikyo Univ., Utsunomiya, 320-8551 Japan\\
        E-mail: \email{furui@umb.teikyo-u.ac.jp}}

\abstract{A new gauge fixing that imposes the U(1) real and quaternion real condition on the correlator of domain wall fermion on the left wall and right wall is proposed.  The method is applied to the calculation of the mass function, the  QCD effective coupling and the charge form factor of the proton. By assuming topological charge change $|Q|=1$, the charge form factor of the proton is approximately fitted by the dipole form factor.\\
The difference of the Kugo-Ojima color confinement parameter in quenched simulation and in unquenched simulation is expected  to be due to an anomaly caused by the vertex of two fermion zero modes and the Bose ghost/gluon.
 }

\FullConference{The XXVII International Symposium on Lattice Field Theory\\
July 26-31, 2009\\
Peking University, Beijing, China}

\begin{document}

\section{Introduction}
The infrared (IR) QCD is characterized by the color confinement and the chiral symmetry breaking.   We performed \cite{SF08b} lattice simulations of the domain wall fermion which preserves the chiral symmetry in the zero-mass limit using the full QCD  gauge configurations of the RBC/UKQCD collaboration of $16^3\times 32\times 16$ lattice \cite{AABB07} and compared with results of staggered fermion of the MILC collaboration \cite{MILC01}.  In these lattice simulations and in comparison with other works, we observed qualitative differences between quenched and unquenched simulations in the QCD effective running coupling, and in the  Kugo-Ojima  parameter \cite{KO79} which is calculated numerically by the inverse of the Faddeev-Popov operator whose zero-modes subtracted at the zero momentum \cite{FN03,FN04,FN05,SF08a}. 
The difference of the quenched and the unquenched simulation could be attributed to the fermion zero mode. In finite temperature hamiltonian QCD, zero-modes are localized around monopoles whose position is defined by the boundary condition on the time axis \cite{vB08}.

In Sect.2, we review the IR QCD and Gribov-Zwanziger-Kugo-Ojima theory, and
in Sect.3, we show the method of gauge fixing using the boundary condition of fermions on the domain wall. In Sect.4, the quark-gluon coupling is presented and in Sect.5, the charge form factor of the proton is presented. Discussion and comments on the Kugo-Ojima parameter are given in Sect.6.

\section{The IR QCD and Gribov-Zwanziger-Kugo-Ojima theory}
The Landau gauge QCD or the Coulomb gauge QCD in the IR region suffers from the problem of gauge uniqueness. Gribov \cite{Gr78} defined the so called Gribov region $\Omega$, which is convex and bounded in every direction and every gauge orbit passes through it at least once. Zwanziger \cite{Zw89} showed that the cut-off of the functional integral at the Gribov horizon $\partial \Omega$ may be replaced by the Boltzmann weight. He defined the free field action $S_0[A]$ and an ellipsoid in $A$ space $Q[A]$  and replaced the measure inside the Gribov horizon\[
d\,A exp(-S_0)\delta(c-Q[A])\to d\mu_\gamma=d\,A exp(-S_0)exp(-\gamma VQ[A])
\]
where $\gamma$ is defined from the self-consistency $\langle Q[A]\rangle_\gamma=c$, where the l.h.s is measured with use of $d\mu_\gamma$. When $\gamma\ne 0$, there appears in addition to the Fermi ghost $c$ whose correlator $\langle c\bar c\rangle$ yields the Faddeev-Popov determinant, but the Bose ghost $\phi$ which couples to the quark and yields the linearing rising potential \cite{Zw09}.

Kugo and Ojima \cite{KO79} assumed $\gamma=0$ and the BRST symmetry in the IR region and defined the color confinement parameter $u(k^2)$ as
\[
\langle D_\mu c^a(A_\nu\times c)^b\rangle_k=(\delta_{\mu\nu}-\frac{k_\mu k_\nu}{k^2})\delta^{ab} u(k^2),
\]
and $u(0)=-1$ as the color confinement condition.
In this theory, the ghost dressing function $F(k)$ is expressed as
\[
F(k)=\frac{1}{1+u(k^2)+o(k^2)}
\] 
and the ghost propagator should be infrared divergent.

Quenched lattice simulations indicate, however, $u(0)$ is about -0.8 and unquenched lattice simulation indicates that $u(0)\sim -1$ \cite{FN04}, but the ghost propagator does not show the infrared divergence.
Recently, violation of the BRST symmetry and infrared finiteness of the ghost propagator are proven \cite{Dudal09, Kondo09a, Kondo09b}, but whether $u(0)$ becomes -1 in unquenched simulation is not clear.

In perturbative QCD (pQCD), the ghost propagator is usually calculated via Fermi ghost-gluon coupling, but when $\gamma\ne 0$ there is a coupling of gluon to Bose ghost $\phi$ which couples to quarks. The fermionic zero mode should modify the measure $d\mu$ and we want to evaluate the zero mode effects using a specific gauge fixing.

\DOUBLEFIGURE[h] {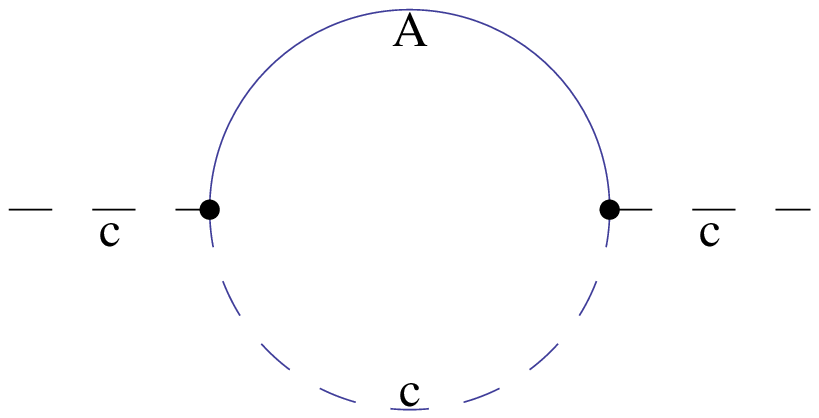,width=4cm}{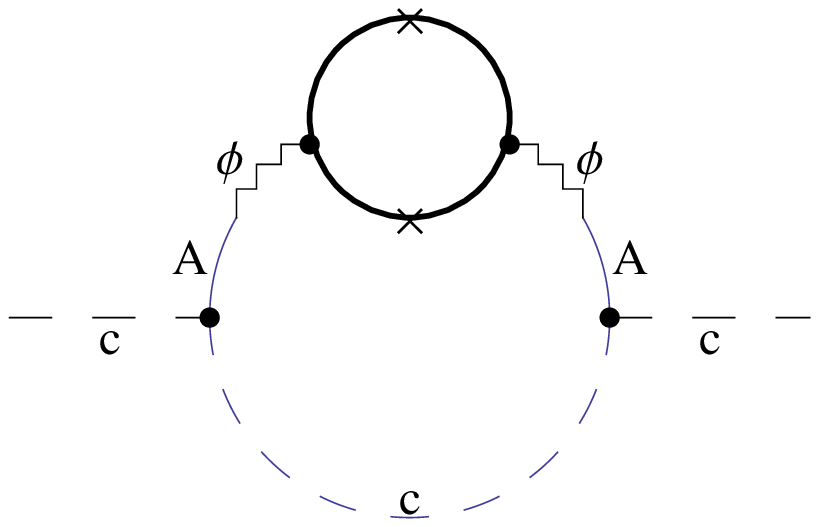,width=4.5cm}{The ghost-gluon loop contribution to the ghost propagator. Solid curves are gluons and dashed lines are Fermi ghosts.}{The quark loop contribution in the ghost propagator. Thick solid curves are quarks, zig-zag lines are Bose ghosts. }

\section{The gauge fixing using the boundary condition of fermions on the domain wall}

On the lattice, the expectation value of the quark propagator $S(p)$ consists of spin dependent $\mathcal{A}p$ part and 
spin independent ${\mathcal B}$ part. The propagator is written as
\begin{equation}
S(p)=\frac{-i{\mathcal A}p+{\mathcal B}}{{\mathcal A} (p^2+{\mathcal{M M}}^\dagger)}
=\frac{ -i{\mathcal A} p+{\mathcal B}}{{\mathcal A} p^2+{\mathcal{M B}}^\dagger}.
\end{equation}
We specify the propagator of the left-handed quark by the suffix $L$  and the right-handed quark by the suffix $R$. With use of the renormarization factor $Z_A$  and $Z_B$, they are defined as

\[
{\rm Tr} \langle \bar\chi(p,s) P_{L/R} \Psi(p,s)\rangle=Z_B(p)(2N_c) {\mathcal B}_{L/R}(p,s),
\]
and
\[
{\rm Tr} \langle \bar\chi(p,s) i\Slash{p}P_{L/R}\Psi(p,s)\rangle=Z_A(p)/(2N_c) i{\bf p}{\mathcal A}_{L/R}(p,s),
\]
where $\displaystyle p_i=\frac{1}{a}\sin \frac{2\pi \bar p_i}{n_i}$ ($\bar p_i=0,1,2,\cdots,n_i/4$).
The mass is defined as
\[
{\mathcal M}(\hat p)=\frac{Re[{\mathcal B}_R(p,L_s/2)]}{Re[{\mathcal A}_R(p,L_s/2)]} \quad {\rm and} \quad 
{\mathcal M}^\dagger(\hat p)=\frac{Re[{\mathcal B}_L(p,L_s/2)]}{Re[{\mathcal A}_L(p,L_s/2)]}.
\]

\subsection{The U(1) real condition}
 Since we adopt the representation of fermion spinors such that the $\gamma_5$ is diagonal, the fermion field $\psi$ can be gauge transformed in the fifth dimension as
\begin{equation}
\psi\to e^{i\eta\gamma_5}\psi, \qquad \bar\psi \to \bar\psi e^{-i\eta\gamma 5},
\end{equation}
such that on the left and on the right domain wall, the propagator becomes approximately real by choosing \cite{SF08b}
\[
|e^{i\theta L}e^{i\eta}-1|^2+|e^{i\theta R}e^{-i\eta}-1|^2
\]
be minimum.

We sample wise measure the propagator by using the transformation matrix from the left boundary
\[
a_x^0\sigma_1+a_y^0\sigma_2+a_z^0\sigma_3+a_t^0\,i I=\left(\begin{array}{cc}a_{0}& b_{0}\\
                       c_{0}& d_{0}\end{array}\right)
\]
which is given by the sample average of color diagonal components 
to the right boundary
\[
a_x^{L_s-1}\sigma_1+a_y^{L_s-1}\sigma_2+a_z^{L_s-1}\sigma_3+a_t^{L_s-1}\,i I=\left(\begin{array}{cc}a_{L_s-1}& b_{L_s-1}\\
                       c_{L_s-1}& d_{L_s-1}\end{array}\right)
\]

\subsection{The quaternion real condition}

As the transformation matrix of the fermion at $s=0$ to $s=L_s-1$, we adopt the ansatz \cite{CG81}, 
\begin{eqnarray}
g_0&=&\left(\begin{array}{cc}e^{-\nu}&0\\
                           0&e^\nu\end{array}\right)
\left(\begin{array}{cc}\zeta^1&\rho\\
                       0&\zeta^{-1}\end{array}\right)
\left(\begin{array}{cc}e^{\mu}&0\\
                           0&e^{-\mu}\end{array}\right)\nonumber\\
&=&\left(\begin{array}{cc}e^\gamma\zeta^1&f(\gamma,\zeta)\\
                       0&e^{-\gamma}\zeta^{-1}\end{array}\right).
\end{eqnarray}
In our five-dimensional domain wall fermion case, $\gamma=\mu-\nu$ and $\mu, \nu$ contain the phase in the fifth direction $i\eta$. 
\begin{equation}
2\mu=i\omega_2/\pi_2-i\eta=(p_x+ip_y)\zeta+ip_t-p_z-i\eta
\end{equation}
\begin{equation}
2\nu=i\omega_1/\pi_1+i\eta=(p_1x-ip_y)\zeta+ip_t+p_z+i\eta
\end{equation}
 The quaternion reality condition of the transformation matrix $g(\gamma,\zeta)$ gives
\begin{equation}
\left(\begin{array}{cc}a_{L_s-1}& b_{L_s-1}\\
                       c_{L_s-1}& d_{L_s-1}\end{array}\right)
\left(\begin{array}{cc}\zeta^{1}e^\gamma& f\\
                       0&\zeta^{-1}e^{-\gamma}\end{array}\right)
=\left(\begin{array}{cc}\zeta^{1}e^{-\gamma}& \bar f\\
                       0&\zeta^{-1}e^{\gamma}\end{array}\right)
\left(\begin{array}{cc}a_{0}& b_{0}\\
                       c_{0}& d_{0}\end{array}\right)
\end{equation}
where $\displaystyle \bar f=\overline{f(\bar\gamma,-\frac{1}{\bar\zeta})}$.

The function $f$ and $\bar f$ taken in \cite{CG81} is.
\[
f=\frac{d_0 e^\gamma-\frac{1}{a_{L_s-1}}e^{-\gamma}}{\psi},\quad 
\bar f=\frac{\frac{1}{d_{L_s-1}} e^\gamma-a_0e^{-\gamma}}{\psi}.
\]

In general $c_0$ and $c_{L_s-1}$ are polynomials of $\zeta$, and they satisfy $c_{L_s-1}\zeta^{1}=c_0 \zeta^{-1}$.  We define
\[
\psi=\hat c_{-1}\zeta^{-1}+\hat c_{1}\zeta^{1}+\delta
\]
where $\hat c_{-1}=c_{L_s-1}$ and $\hat c_{1}=c_0$ and $\delta$ is a constant, which is defined later. 

The difference
\begin{equation}
\Delta L/R=\left(\begin{array}{cc}a_{L_s-1}& b_{L_s-1}\\
                       c_{L_s-1}& d_{L_s-1}\end{array}\right)_{L/R}
\left(\begin{array}{cc}\zeta^{1}e^\gamma& f\\
                       0&\zeta^{-1}e^{-\gamma}\end{array}\right)
-\left(\begin{array}{cc}\zeta^{1}e^{-\gamma}& \bar f\\
                       0&\zeta^{-1}e^{\gamma}\end{array}\right)
\left(\begin{array}{cc}a_{0}& b_{0}\\
                       c_{0}& d_{0}\end{array}\right)_{L/R}
\end{equation}
should be small, if the left wall and the right wall are correlated by the self-dual gauge transformation.

We choose $\zeta^{1}=\sqrt{\frac{c_0}{c_{L_s-1}}}$ to make $(\Delta L/R)_{21}$ consistent with 0 and obtain $e^\gamma$ and $\delta$, which approximately satisfy the quaternion reality. 
Corresponding to the sample-wise selection of larger absolute value of $\delta$, i.e. a large zero-mode component, we assign a parameter $ind=1$ for larger and 2 for smaller, and multiply the phase $(-1)^{ind}$ to the expectation value of the quark wave function that is used in the calculation of the effective mass of the quark. 
This phase factor can be absorbed in the phase $e^{i\eta}$ introduced to approximately satisfy the U(1) real condition.

\section{The quark-gluon coupling}

In Fig.1, the quark-gluon coupling in the Coulomb gauge of the domain wall fermion is compared with that of the staggered fermion.
Red points below 1.2GeV are the experimental data of the JLab collaboration and the dark point is our result of momenta $q=0.7$GeV/c. The dash-dotted line is the two loop pQCD result and the dashed line is the pQCD with $A^2$ condensates effect derived from the fitting the running coupling \cite{FN05} calculated by using the gauge configuration of full-QCD staggered fermion of the MILC collaboration \cite{MILC01}.

\DOUBLEFIGURE[h]{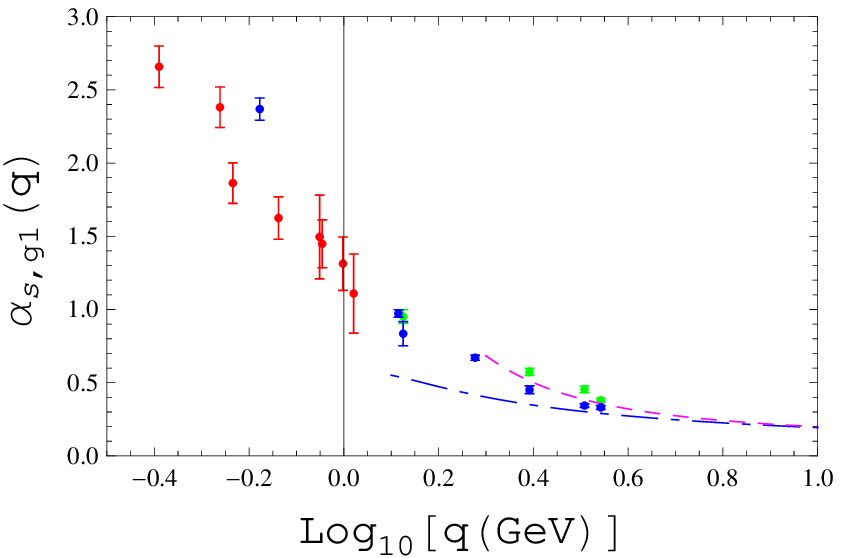,width=7cm} {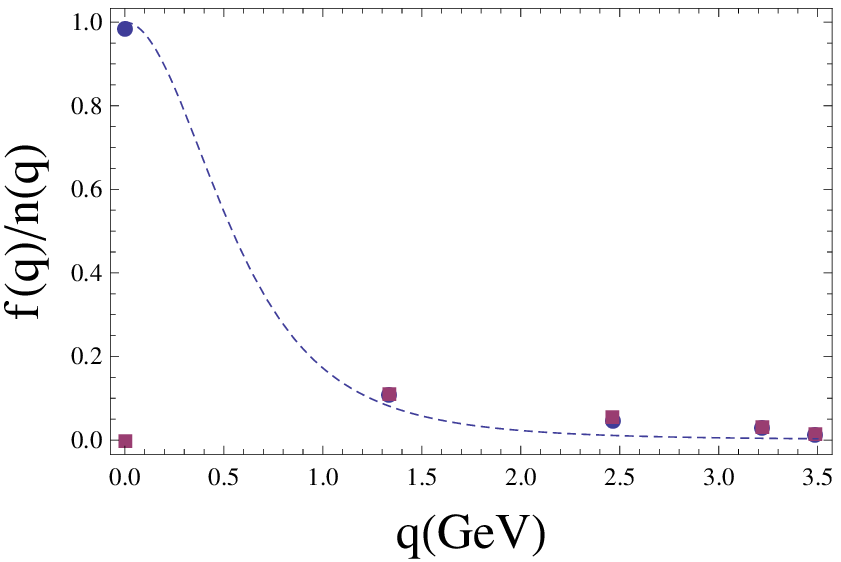,width=7cm}{The effective coupling $\alpha_{s,g_1}(q)$ of domain wall fermion and the gluon without reality selection (greenjand with reality correction(magenta)  in cylinder cut region. Dark low momentum points are that with reality correction.   }{The charge form factor of the proton. Dashed line is the dipole fit of $1/(1+q^2/M^2)^2$ with $M^2=0.71$GeV$^2$.}

\section{The charge form factor of a proton}

We measure the charge form factor of a proton as a ratio of the three point function and the two-point function in momentum space.
We calculate expectation values of $G^{NVN}_{c+}$ etc by diagonalizing the
matrices whose coefficients of $\sigma_1, \sigma_2$ and $\sigma_3$ are real, and then evaluate e.g. 
$\langle (u^a_+C\gamma_5 d^b_-)u^c_+|\gamma_4 \gamma_\mu|(u^a_*C\gamma_5 d^b_*)u^c_*\rangle$ as
\begin{equation}
\frac{{G^{NVN}}_{c+}(p)}{G^{NN}_{c+}(p)}
+\frac{{G^{NVN}}_{a-}(p)}{G^{NN}_{a-}(p)}+\frac{{G^{NVN}}_{b-}(p)}{{G^{NN}}_{b-}(p)}
\end{equation}

The three point function $G^{NVN}_{c+}(p)$ is a product of 
\[
\sum_{a,b,c,c'}S(c_+,c_{+/-},p)\langle V(c_{+/-},c'_{-/+},p)\rangle\gamma_5 S(c'_+c'_{-/+},p)^\dagger\gamma_5 {\bar S}(a,b,p).
\] 
where $\bar S(a,b,p)$ is the product of the propagator of the quark $a$ and that of $b$.
\[
{\bar S}(a,b,p)=S(a_-,a_-,p)\gamma_5 S(b_-,b_-,p)^\dagger\gamma_5
\]
In the two point function $G^{NN}_{c+}(p)$, the factor before ${\bar S}(a,b,p)$ is replaced by the average of $S(c_+,c_{+/-},p)$ and $S(c'_+c'_{-/+},p)^\dagger$.

In the product of $S(c_+,c_{+/-},p) \gamma_5 S(c'_+c'_{-/+},p)^\dagger\gamma_5$ we assume contribution of self-dual field such that one propagator is that of left-handed and the other is right-handed.   It means that we consider a change of the topological charge of one unit in the propagator.  Of course it is an assumption which should be verified by comparison with experimental data.

The sum of the left-handed and the right-handed contributions are to be compared with the vector current form factor.
The form factor at $p=0$ is dominated by the $V(c_{+/-},c'_{-/+},p)$ of the left-handed fermion, while at finite $p$, the left-handed and the right-handed fermion contribute equally.  The normalization at $p=0$ is $1/36$ corresponding to $(3 \,{\rm colors} \times 2 \,{\rm spin \, projections})^2$ , and relative normalization at $p\ne 0$ in the figure is multiplied by 2. (The sum of left-handed and right handed are close to the twice of the left-handed.)

\section{Discussion and comments on the Kugo-Ojima parameter}
In the quenched calculation, whether the $u(0)$ becomes -1 or not depends on the treatment of the total derivative term \cite{Kondo09b}. On the lattice, however, this term is irrelevant. We think the difference of \cite{Dudal09} and \cite{Kondo09b} exists in another kind of total derivative term.   When the quark field contribution in the quark-quark-Bose ghost/gluon($\phi/A$) vertex \cite{Zw09} 
\[
g^3 \lambda^a\gamma_\mu S_{q\bar q}\gamma_\nu\lambda^b D_{AA,\mu\lambda} D_{A\phi,\nu\kappa}^{bce}f^{acd}k_\lambda
\]
is replaced in the zero-mass limit by the product $(U^\dagger\partial_\mu U)(U^\dagger\partial_\nu U)$ where $U$ are the quaternions of the fermion zero-modes, it becomes reminiscent of the anomalous current given in \cite{Gr87}, which is expected to stabilize the IR fluctuation: 
\[
j_\kappa^e=-\frac{\epsilon^2}{32\pi^3}\int\frac{d\Omega}{(\epsilon z)^2} d^{abe}F^a_{\kappa\nu}(x)A^b_\nu(x,z)
\]
where $\epsilon_\kappa$ is an arbitrary vector. The anomaly would be cancelled by that in the Baryon sector \cite{tH}.

We think that the contribution of the Bose ghost that couples to quark zero-modes cannot be ignored in the unquenched Kugo-Ojima parameter calculation, and that the Nature of IR QCD possesses a kind of supersymmetry. 

\vskip 0.2 true cm
{\small The author thanks Kei-ichi Kondo for valuable informations. The numerical simulation was performed on Hitachi-SR11000 at High Energy Accelerator Research Organization(KEK) under a support of its Large Scale Simulation Program (No.07-04 and 08-01), and on NEC-SX8 at Yukawa institute of theoretical physics of Kyoto University and at CMC of Osaka university.}

\end{document}